\documentstyle[12pt,twoside]{article}

\begin{document}
\baselineskip 28pt
\begin{flushright}
CRPS-93-31
\end{flushright}
%\maketitle
%\def\quote{\list{}{\rightmargin 3pc \leftmargin 3pc}\item[]}
%\def\quot{\list{}{\rightmargin 5pc \leftmargin 5pc}\item[]}
\begin{center}
\begin{Large}
\begin{bf}
Comment on ``Magnetic Susceptibility of the\\
two-dimensional Hubbard model '' \\
\end{bf}
\end{Large}
\begin{large}
\vspace{0.3cm}
Liang Chen and  A.-M.S. Tremblay \\
D\'epartement de physique et \\
Centre de recherche en physique du solide,\\
Universit\'e de Sherbrooke\\
Sherbrooke, Qu\'ebec, Canada, J1K 2R1\\
\vspace{0.5cm}
\end{large}
\end{center}
\vspace{0.5cm}
%\begin{quote}
\begin{center}
{\large ABSTRACT}
\end{center}
%\begin{small}
The observed magnetic spin susceptibility of high-temperature superconductors
such as La$_{2- x}$Sr$_x$CuO$_4$ increases when $x$ increases from zero, i.e.
as one dopes away from half-filling.  Recent Monte Carlo simulations of A.
Moreo (Phys. Rev. B {\bf 48}, 3380 (1993)) suggest that this behavior can be
reproduced by the two-dimensional Hubbard model only at large coupling, namely,
$U/t$ of order $10$.  Using longer runs, our Monte Carlo simulations show that
the same behavior as for $U/t=10$ is obtained even in the intermediate coupling
regime ($U/t=4$), as long as the temperature is low enough ($T=t/6$) that
strong antiferromagnetic correlations are building up at half-filling.  These
results are consistent with the fact that in two-dimensions, the GRPA should
fail in the parameter range where it predicts a magnetic phase transition.

\vspace{0.2cm}

\noindent P.A.C.S. Numbers: 75.30.Cr, 74.72.-h, 75.10.Lp, 74.20.-z
\\
%\noindent 93-09-06
%\end{small}
%\end{quote}

\newpage
%\vspace{0.2cm}

One of the main puzzles in establishing the applicability of the one-band
Hubbard model to high-temperature superconductors is the behavior of the
uniform magnetic spin susceptibility as obtained, say, by Knight-shift
measurements.$^1$  While the maximum as a function of temperature is easy to
understand qualitatively from the behavior of the Hubbard model at
half-filling in the strong-coupling (Heisenberg) regime, the behavior away
from half-filling and in the intermediate- or weak-coupling regime is
theoretically much more uncertain.  The limiting low-temperature value of the
magnetic susceptibility experimentally increases as a function of doping, a
behavior which at first glance can be reproduced within perturbative approaches
only in models which include second-neighbor hopping$^2$.  With only
nearest-neighbor hopping, non-perturbative effects, such as pseudo-gaps$^3$ are
usually invoked.

To numerically check under what conditions the one-band {\it nearest-neighbor}
Hubbard model reproduces the experimentally observed behavior$^1$, Moreo$^4$
has studied the magnetic spin susceptibility using determinantal quantum Monte
Carlo techniques$^{5,6}$.  She has shown that on a $4\times 4$ lattice at
inverse temperature $\beta = 4/t$ the susceptibility has a maximum at
half-filling and decreases with doping in the intermediate-coupling regime
$U/t=4$, while in the strong-coupling regime $U/t=10$, the susceptibility
increases with hole doping near half filling, resembling the experimental
results$^1$ obtained for La$_{2-x}$Sr$_x$CuO$_4$.  She suggests that
experimental results can be reproduced only when the coupling is quite large,
namely $U/t=10$ or more.

We have used the same quantum Monte Carlo technique on a $4\times 4$ lattice
for $U/t=4$ but at lower temperature, namely $k_BT=t/6$, or inverse temperature
$\beta=6/t$. As shown in Fig. 1, we find that at this lower temperature, the
susceptibility has qualitatively the same behavior as that observed for larger
coupling ($U/t=10$) and larger temperature ($\beta = 4/t$ ) by Moreo.
We measured the structure factor
instead of the uniform magnetic susceptibility but these quantities are easily
related using the fluctuation-dissipation theorem.  To obtain the results of
Fig. 1, one must perform much longer runs than is usually done.  Typically, to
obtain one point on the curve we have used  $\Delta\tau=1/15$ and 100,000
warm-up sweeps of the whole space-time lattice before the measurements were
taken.  Measuring the static structure factor after every update of the lattice
in space, and doing no measurement every other update of the whole space-time
lattice,  we have made of order 10,000,000 measurements, using blocks of 10,000
measurements to eliminate the effect of sticking $^7$ on the estimation of the
statistical error. It takes almost one week of computing time to obtain one
point in Fig.1 with a processor executing 27 million floating-point operations
per second.$^8$

\begin{figure}[htbp]
\vskip 8truecm
\includegraphics{spsus.ps}
\caption{The static magnetic spin susceptibility $\chi_s$ multiplied by
temperature T is plotted as a function of band filling $<n>$.   The units are
$t=1$, $k_B=1$, $\hbar=2$.  Monte Carlo results are shown by points and error
bars.  The solid line is a guide to the eye. The dotted line shows the result
for the free case (U=0) for the same system size and temperature.}
\end{figure}

Our results and those of Moreo$^4$ suggest that it is the presence of strong
antiferromagnetic correlations at half-filling which yield a maximum in the
magnetic susceptibility away from half-filling.  Fig.  3 of Ref. 6 shows that
at $U/t=4$ on a $4\times 4$ lattice, the antiferromagnetic correlation length
at half-filling has basically reached the size of the lattice$^9$ at the
temperature we studied ($\beta=6/t$), while it has not quite made it at the
temperature ($\beta=4/t$) studied by Moreo$^4$.  Evidently, at larger values of
$U/t$, the antiferromagnetic correlation length could reach the system size
even at $\beta=4/t$.  The results of Monte Carlo simulations on a related model
also suggest the key role of strong antiferromagnetic correlations at
half-filling: In the three-band model, Dopf et al.$^{10}$ have found a
parameter range where the behavior observed experimentally is reproduced
qualitatively and it seems that, in this parameter range, strong
antiferromagnetic correlations are also present at half-filling.$^{11}$

Whether these results of simulations on finite lattices have anything to do
with the infinite system is a tricky question.  A zeroth-order check consists
in making sure that the $U=0$ result has the same qualitative dependence on
filling for both the finite and the infinite lattice.  In Fig.1, we plotted the
$U=0$ result for a $4\times 4$ lattice with a dotted line.  Clearly, the spin
susceptibility decreases as the system is doped away from half-filling, in
qualitative agreement with the infinite system.  This kind of agreement is not
true at all fillings since on a finite lattice the spin susceptibility is not a
monotonous function of filling while it is for the infinite system.  Moreo$^4$
has argued that finite-size effects should be smaller for larger interaction
strengths because of localization effects.  This is almost certainly true,
unless interactions introduce some other effect, such as a phase transition.
We can guess what should happen at $U/t=4$ by using some analytical results.
It has been shown$^{12}$ that the magnetic structure factor obtained by Monte
Carlo simulations is well described, not too close to half-filling, by the
Generalized Random Phase Approximation (GRPA), if one takes into account
two-particle correlations (Kanamori-Brueckner screening) by renormalizing the
value of $U$.  Using the renormalized value of $U_{rn}/t=2.2$ appropriate for
the bare value $U/t=4$, we find that long-range order sets in for a filling
about equal to that where the maximum appears in Fig.1.  This is consistent
with the fact that in two-dimensions, the GRPA should start to fail when it
predicts a magnetic phase transition.  Indeed, a better approximation would
take into account Mermin-Wagner fluctuations which prohibit long-range order in
two-dimensions.  How to include these fluctuations in itinerant electron
theories is a problem which cannot presently be answered by Monte Carlo
simulations but which is beginning to be successfully addressed
analytically.$^{13}$  It would be reasonable to expect that in the regime where
the mean-field phase transition found using the GRPA is suppressed by thermal
fluctuations ($T<T^{(c)}_{GRPA}$) the large antiferromagnetic fluctuations will
nevertheless decrease the uniform spin susceptibility (at half-filling in
Fig.1, it is smaller than the non-interacting value.).  Because the tendency
towards antiferromagnetism increases towards half-filling, the infinite-size
two-dimensional Hubbard model would then show the same kind of behavior as the
finite-size system of Fig.1.$^{14}$  Consistency with experiment might then
occur not only in the strong-coupling limit, but also at relatively smaller
couplings as long as $T^{(c)}_{GRPA}$ is not small compared with experimentally
studied temperatures.  In both the present work and that of Ref.4 however, the
system sizes are much too small to yield a definitive answer to the question of
the infinite-size limit.

\vspace{0.3cm}
\begin{description}
\item[] \hspace{1.5cm} {\large\bf Acknowledgments}
\end{description}

We would like to thank the A. Moreo for sending us a preprint of her work$^4$
before publication.  We are also grateful to her for sharing her data, and for
useful comments on the manuscript.  We also thank Y.Vilk and A.Veilleux for
many insightful comments.  We gratefully acknowledge the support of the Natural
Sciences and Engineering Research Council of Canada (NSERC), the Fonds pour la
Formation de Chercheurs et l'Aide \`a la Recherche (FCAR) of the Government of
Qu\'ebec, and (A.-M.S.T.) the Canadian Institute of Advanced Research
(C.I.A.R.) and the Killam Foundation.

%\vspace{0.3cm}
\newpage

\begin{center}
{\large\bf References}
\end{center}
\begin{description}
\item[1] W.W. Warren et al, Phys. Rev. Lett {\bf 62}, 1193 (1989); R.E.
Waldstedt et al., Phys.  Rev. B {\bf 41}, 9574 (1990);  R.E. Waldstedt et
al., Phys. Rev. B {\bf 44}, 7760 (1991); M.  Takigawa et al., Phys. Rev. B
{\bf 43}, 247 (1991); H. Alloul et al., Phys. Rev. Lett. {\bf 63}, 1700
(1989).   J.B. Torrance et al., Phys. Rev. B {\bf 40}, 8872 (1991).
\item[2] Pierre B\'enard, Liang Chen, and A.-M.S. Tremblay, Phys. Rev B {\bf
47}, 15,217 (1993); K. Levin, J.H. Kim, J.P. Lu and Q. Si, Physica C {\bf
175}, 449 (1991).
\item[3] A.P. Kampf, and J.R. Schrieffer, Phys. Rev. B {\bf 42}, 7967 (1990).
\item[4]A. Moreo, Phys. Rev. B {\bf 48}, 3380 (1993).
\item[5] R. Blankenbecler, D.J. Scalapino, and R.L. Sugar, Phys.  Rev.  D {\bf
24}, 2278 (1981): J.E.  Hirsch, Phys. Rev. B {\bf 31}, 4403 (1985);  For a
review, see E.Y. Loh, et J.E.  Gubernatis, in {\it Electron Phase Transitions}
p.  177-235, edited by W.  Hauke, Y.V. Kopaev (Elsevier, Amsterdam, 1992).
\item[6] S.R. White, D.J. Scalapino, R.L. Sugar, E.Y. Loh,  J.E.  Gubernatis,
R.T. Scalettar, Phys. Rev. B40, 506 (1989).
\item[7] Liang Chen and A.-M.S. Tremblay, Int. J. Mod. Phys. B {\bf 6},547
(1992).
\item[8] Following J.E. Hirsch, Phys. Rev. B {\bf 35}, 1851 (1987), we used the
transverse correlations as the estimator.  As pointed out to us by Moreo
however, it turns out that for this problem the longitudinal estimator is
less noisy.  We have checked that it is indeed possible to obtain the
same results with the longitudinal estimator using runs which are five to ten
times shorter.
\item[9] When the magnetic structure factor starts to scale with system size
(as it does on the plateaus in Fig. 3 of Ref. 6),
this indicates that the correlation-length has reached the system size.  This
is why we take proximity to the
plateau as an indication that the antiferromagnetic fluctuations are becoming
strong.
We are not interested
in the zero-temperature limit per se.
\item[10] G. Dopf, A. Muramatsu, and W. Hanke, Phys. Rev. Lett. {\bf 68}, 353
(1992). See Fig. 3 of this paper.
\item[11] G. Dopf, A. Muramatsu, and W. Hanke, Phys. Rev. B {\bf 41}, 9264
(1990). See Fig. 8 of this paper.
\item[12] Liang Chen, C. Bourbonnais, T.C. Li, and A.-M.S. Tremblay, Phys.
Rev. Lett. {\bf 66}, 369 (1991); N. Bulut, D.J. Scalapino, and S.R. White,
Phys. Rev. B {\bf 47}, 2742 (1993).
\item[13] Y. Vilk (unpublished).
\item[14] Millis and Monien have recently argued in favor of such a
spin-density-wave explanation of the uniform magnetic susceptibility in La$_{2-
x}$Sr$_x$CuO$_4$: A.J. Millis, and H. Monien, Phys. Rev. Lett. {\bf 70}, 2810
(1993).

\end{description}

%\newpage
%\begin{center}
%{\bf Figure Caption}
%\end{center}
%\vspace{1.cm}
%Fig. 1 The static magnetic spin susceptibility $\chi_s$ multiplied by
%temperature T is plotted as a function of band filling $<n>$.   The units are
%$t=1$, $k_B=1$, $\hbar=2$.  Monte Carlo results are shown by points and error
%bars.  The solid line is a guide to the eye. The dotted line shows the result
%for the free case (U=0) for the same system size and temperature.
%\begin{enumerate}
%\end{enumerate}

\end{document}